\documentclass{article}

\usepackage[main,final]{neurips_2025}

\usepackage[utf8]{inputenc}
\usepackage[T1]{fontenc}
\usepackage{hyperref}
\usepackage{url}
\usepackage{booktabs}
\usepackage{amsfonts}
\usepackage{amsmath}
\usepackage{amssymb}
\usepackage{amsthm}
\usepackage{nicefrac}
\usepackage{microtype}
\usepackage{xcolor}
\usepackage{graphicx}

\title{Normalizing Flows to Reconstruct Pseudo-PDFs}

\author{Anonymous Authors}

\author{Yamil Cahuana Medrano\\
  Department of Physics\\
  William \& Mary\\
  Williamsburg, VA 23185 \\
  \texttt{yacahuanamedra@wm.edu} \\
  \And
  Kostas Orginos \\
  Department of Physics \\
  William \& Mary \\
  Williamsburg, VA 23185 \\
  \texttt{knorgi@wm.edu}
}
\begin{document}

\maketitle

\begin{abstract}

We investigate a normalizing-flow approach for reconstructing parton distribution functions (PDFs) from synthetic matrix-element data. Our framework combines Gaussian Process priors with invertible neural networks to learn a posterior distribution over PDFs consistent with limited Ioffe-time data. We demonstrate that the architecture preserves physical constraints and extrapolation properties.
\end{abstract}

\section{Introduction}

Parton distribution functions (PDFs) are essential to our understanding of hadron structure, representing a central pillar of particle physics research. Despite their importance, direct calculation remains challenging due to the non-perturbative nature of quark and gluon interactions at low energy scales. In the recent decade, several lattice frameworks \cite{Ji:2013dva, Radyushkin:2017cyf} have been proposed to investigate these quantities by computing non-local matrix elements that follow

\begin{equation}
\langle P | \bar{\psi}(z)\gamma^\mu W(z,0)\psi(0) | P \rangle=P^\mu M(\nu,z^2)+z^\mu N(\nu,z^2)
\end{equation}

where $P$ and $z$ represent the momentum of the hadron and the space-like separation, respectively, and $\nu=P\cdot z$ is the Lorentz-invariant quantity called the Ioffe-time. Using the ratio scheme to remove UV divergences, the pseudo-PDF $P(x,z^2)$ can be obtained through the solution of $M(\nu,z^2)=\int_{-1}^1 dx e^{i\nu x} P(x,z^2)$, or equivalently
\begin{equation}\label{eq:cossin}
    \operatorname{Re} M(\nu,z^2) = \int_{0}^1 dx \cos(\nu x) q_{-}(x,z^2),\qquad \operatorname{Im} M(\nu,z^2) = \int_{0}^1 dx \sin(\nu x) q_+(x,z^2),
\end{equation}
where $q_+(x,z^2) = P(x,z^2) - P(-x,z^2)$ represents the singlet (C-even) distribution, and $q_-(x,z^2) = P(x,z^2) + P(-x,z^2)$ represents the valence (C-odd) distribution. From this point we will not consider the z dependence of PDF, because the matching procedure does not interfere with the inverse problem at hand.

Gaussian processes (GP) allow one to encode smoothness, correlations, and physically motivated constraints within a non-parametric Bayesian setting \cite{Rasmussen2006Gaussian}, while preserving a controlled description of uncertainty from simple regression problems to the solution of inverse problems in the context of differential equations \cite{Raissi_2017}.
 Several approaches have been explored to address this inverse problem,   
 including Bayesian methods~\cite{Karpie_2019}, neural-network            
 parameterizations~\cite{Del_Debbio_2021, Cichy_2019}, and Gaussian       
 process regression~\cite{Alexandrou_2020, candido2024bayes,              
 Medrano:2025}, which have been shown to provide robust PDF                
 reconstructions from limited Fourier components in lattice QCD.

In practice, the reconstruction depends on the data and physical constraints, but some precautions have to be taken to constrain the possible behavior in the extrapolation region of Ioffe time, where lattice information is limited or non-existent. Previous work \cite{Dutrieux:2024rem} introduced a physically motivated prescription to fix this prior scale, thereby controlling the uncertainty in the reconstructed distributions. That approach is transparent, efficient, and well justified from the physics of the problem.

The motivation of the present work is to explore if this hyperparameter-selection step can be mapped to a normalizing-flow architecture, called in the literature Invertible Neural Networks, and retain the flexibility of GPs while exploring if the specified calibration procedure is present in this alternative method and leads to equivalent results.

\section{Method}
Invertible Neural Networks (INNs)~\cite{ardizzone2019analyzinginverseproblemsinvertible}, a class of normalizing flows~\cite{papamakarios2021norm}, are designed to learn the full posterior distribution of the solution to an inverse problem.
They achieve this by establishing a bijective mapping $f$ between the space of physical quantities of interest and the space of observables. Since these spaces naturally have different dimensions, bijectivity is strictly enforced by introducing latent variables and zero padding. The effectiveness of this architecture relies on three components: structurally invertible affine coupling layers, a bidirectional loss function, and the ability to train on an extensive dataset generated through the forward mapping.

\subparagraph{GP prior.} A PDFs data set can be generated by sampling from a normal distribution. In this context, Gaussian processes (GPs) provide an efficient way to sample functions from a Gaussian distribution over a discretized $x$-grid. This approach allows us to generate the training data
\begin{equation}
    q_{\text{train}}(x) \sim \mathcal{N}\left( \mu(x), K(x,x')\right),
\end{equation}
considering the normalization and the boundary condition
\begin{equation}
    B_0\cdot q=\int_0^1 \delta(x-1) q(x)dx= q(1)=0\qquad B_1\cdot q=\int_0^1 q(x)dx=1.
\end{equation}
We have imposed with Lagrange multipliers $\lambda_{1,2}$ such conditions in covariance and mean as
\begin{equation}
    K = K^{-1}_{\text{log-rbf}} + \frac{B_0^\top B_0}{\lambda_1} + \frac{B_1^\top B_1}{\lambda_2} \quad \mu = K^{-1} \left(K^{-1}_{\text{log-rbf}} \cdot\mu_0 + \frac{B_1^\top}{\lambda_2}\right)
\end{equation}
where $\mu_0(x)=2(1-x)$ and the kernel $K_{\text{log-rbf}}(x,x')=30\cdot e^{-\frac{|\log(x)-log(x')|^2}{0.72}}$. To generate the corresponding Ioffe-time distributions (ITDs), we approximate the integrals in Eq.~\eqref{eq:cossin} using second-order finite elements, yielding a discrete linear operator $B$, such that
\begin{equation}
    [M_{\text{train}}]_{n} = [B]_{n\times k} \cdot [q_{\text{train}}]_{k}.
\end{equation}
with $n = 1,\dots N_\nu$ and $k=1,\dots N_x$.
In our setting, the PDF is discretized on an equally spaced grid of $N_x= 129$ points in ($10^{-3}$, $1-10^{-8}$), while only a limited number of ITD data points, $N_\nu \sim 10$, are available. The solution of the inverse problem consists in the construction of a bijector $f:\mathbb{R}^{N_x} \rightarrow \mathbb{R}^{N_\nu + N_z}$, where $N_z$ denotes the dimension of the latent space such that $N_x = N_\nu + N_z$. Following~\cite{ardizzone2019analyzinginverseproblemsinvertible}, we parameterize the missing information in the noise dimension using a fixed number $N_{\text{noise}}$ of variables, and fill the remaining degrees of freedom with $N_{\text{pad}}$ zeros such that $N_z = N_{\text{noise}} + N_{\text{pad}}$. This procedure guarantees that the INN operates on equal input and output dimensions, while access to a virtually infinite data set ensures convergence of the posterior distribution.
A fundamental step in data processing is normalization. We normalize each sample using the prior mean $\langle q\rangle$ and point-wise variance 
$\mathrm{std}(q)=\sqrt{K_{\text{log-rbf}}(x,x)}$ by 
$q_{train}\rightarrow\frac{q_{\text{train}}-\langle q\rangle}{\mathrm{std}(q)}$. This step helps with stability for the divergent functional form of the PDFs and the convergence of the loss function by addressing the different scales of the PDF in the $x$-grid.

\begin{figure}[t]
  \centering
  \includegraphics[width=0.9\linewidth]{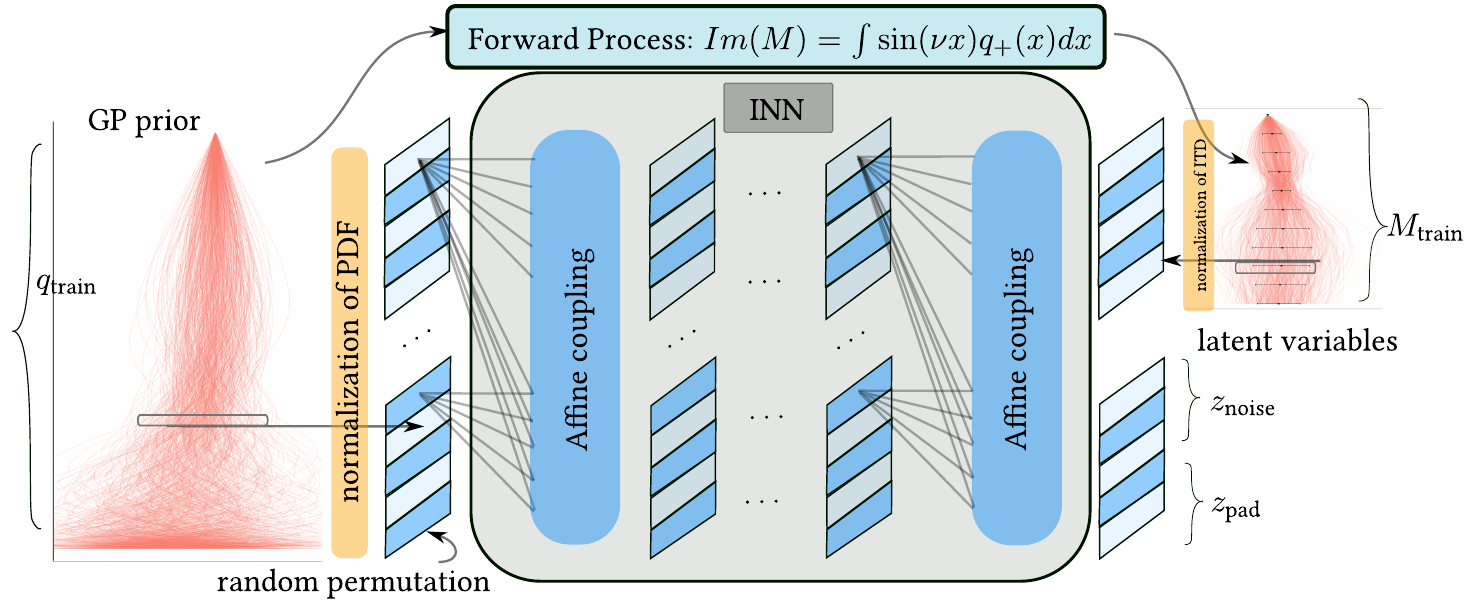}
  \caption{Schematic representation of the INN architecture with forward integration using finite elements for the imaginary part of the matrix elements.}
  \label{fig:test}
\end{figure}

\subparagraph{Affine Coupling Layers.} Our INN is based on affine coupling layers, following the Real NVP architecture \cite{dinh2017densityestimationusingreal,bishop2023learning}. The input vector $q$ is partitioned into two disjoint blocks, $q=[q_1, q_2]$, in a random but fixed manner. To ensure invertibility, the forward transformation from input to output variables $M=[m_1,m_2]$ is defined as
\begin{gather}
    m_1=q_1,\qquad m_2=q_2\odot e^{s(q_1)}+t(q_1).
\end{gather}
where $\odot$ denotes element-wise multiplication, and the scaling $s$ and translation $t$ operations can be arbitrarily complex functions. In this work, $s$ and $t$ are parameterized by standard feed-forward neural networks. By stacking multiple affine coupling layers, we construct the full INN architecture. 
The key advantage of this formulation is that the inverse mapping can be computed analytically and efficiently without needing to invert the neural networks $s$ and $t$. By reversing the sequential operations, the inverse pass is explicitly given by:
\begin{gather}
    q_1 = m_1, \qquad 
    q_2 = \left(m_2 - t(m_1)\right) \odot e^{-s(m_1)}.
\end{gather}
In the inverse mapping, the input $M = [m_1, m_2]$ consists of concatenated ITD data, latent variables, and zero-padding elements. The full INN architecture is constructed by stacking six affine coupling layers $g:\mathbb{R}^{N_x} \rightarrow \mathbb{R}^{N_\nu + N_z}$, such that $f=g^{(1)}\circ \cdots \circ g^{(6)}$, as shown in Fig.~\ref{fig:test}.
Each coupling block contains two subnetworks $s, t$, each a 3-layer MLP with hidden dimension 256 and LeakyReLU(0.2) activations. The scale outputs are clamped via $\sigma(s) = \tanh(s)$ to prevent instabilities.

\subparagraph{Maximum Mean Discrepancy.}
The third component of our architecture is the training objective. To train the INN to accurately reproduce the full posterior, we employ the maximum mean discrepancy (MMD). As shown in \cite{mMMD}, MMD is a kernel-based technique that provides a robust measure of the distance between two probability distributions based on their samples. 

Given two sets of samples, $\{x_i\}_{i=1}^n$ from the target distribution and $\{y_j\}_{j=1}^m$ from the model distribution, the MMD estimator is given by
\begin{equation}
\mathcal{L}_{\mathrm{MMD}^2}(X,Y) =\frac{1}{n^2}\sum_{i=1}^n \sum_{j=1}^n k(x_i,x_j) + \frac{1}{m^2}\sum_{i=1}^m \sum_{j=1}^m k(y_i,y_j) -\frac{2}{nm}\sum_{i=1}^n \sum_{j=1}^m k(x_i,y_j).
\end{equation}

where $k(\cdot, \cdot)$ is a positive-definite kernel. Following \cite{ardizzone2019analyzinginverseproblemsinvertible}, we utilize an Inverse Multi-Quadratic (IMQ) kernel
\begin{equation}
    k_{IMQ}(\mathbf{x},\mathbf{y}) = \frac{1}{1 + \frac{\|\mathbf{x}-\mathbf{y}\|^2}{\sigma}},
\end{equation}
which is effective for comparing distributions with heavy tails while remaining suitable for Gaussian-like samples.
 The active parameter in the kernel is chosen to be $\sigma\in [0.25 s, 0.5 s, s, 2s, 4s]$ with $s=\mathrm{median}\bigl(\{\|x_i-y_j\|^2=\|x_i\|^2 +\|y_j\|^2- 2x_i^\bot y_j\}\bigr)$.

In the context of our inverse problem, this loss is evaluated in both directions to ensure that our neural network learns the distribution of the input and output, comparing the forward-mapped ITD data and latent variables against their prior distributions, and comparing the inverse-mapped predicted PDFs, $q_{\text{INN}} = f^{-1}(M=M_{train}, z_{noise},z_{pad})$, against the target forward-simulated training data, $q_{\text{train}}$. Similarly, the forward INN data is generated by $M_{INN},z^{noise}_{INN},z_{INN}^{pad}=f(q_{train})$.
We optimize the following loss
\begin{gather}
\mathcal{L} = \lambda_x\,\mathcal{L}_{\mathrm{MMD}^2} (q_{\mathrm{INN}}, q_{\mathrm{train}}) +\lambda_z\, \mathcal{L}_{\mathrm{MMD}^2}\! \left((M_{\mathrm{INN}}, z^{noise}_{\mathrm{INN}}),(M_{\mathrm{train}},z_{\mathrm{noise}}) \right) \nonumber\\
+ \lambda_y\,\|M_{\mathrm{INN}}-M_{\mathrm{train}}\|_{1}
+ \lambda_{\mathrm{pad}}\,\|z^{\mathrm{pad}}_{INN}\|_{2}^{2}.
\end{gather}
where $\lambda_x=\lambda_y=\lambda_z=1/3$ and $\lambda_{\mathrm{pad}}=1/6$. 
The norm $\|\cdot\|_{1}$ term is designed to accelerate convergence to the true distribution in the forward direction, and the $\|\cdot\|_{2}$ term minimizes the padded output to zero.
 
The network parameters are optimized using the Adam optimizer~\cite{kingma2017adammethodstochasticoptimization} with an initial learning rate of $10^{-4}$ and a cosine-annealing warm-restart schedule \cite{loshchilov2017sgdrstochasticgradientdescent}. Gradients are clipped to unit norm to improve training stability. All results were obtained from models trained on an NVIDIA A100 GPU. 

\section{Results}
\subparagraph{Closure Test.}
To evaluate the performance of the INN, we design a parametric PDF and map it to the Ioffe-time distribution to generate the closure test. This is done by proposing the following  
\begin{equation}
    q_{val}(x)=\frac{x^\alpha(1-x)^\beta}{B(\alpha+1,\beta+1)}\rightarrow M_{val}=B\cdot q_{val}
\end{equation}
where $\alpha=\mathcal{N}(-0.2,0.5)$ and $\beta=\mathcal{N}(3.0,0.5)$, and $q_{val}$ is discretized in the same grid used for the GP prior. The same normalization procedure with respect to the prior was applied to the ITD data points. 

\begin{figure}[t]
  \centering
  \includegraphics[width=0.95\linewidth]{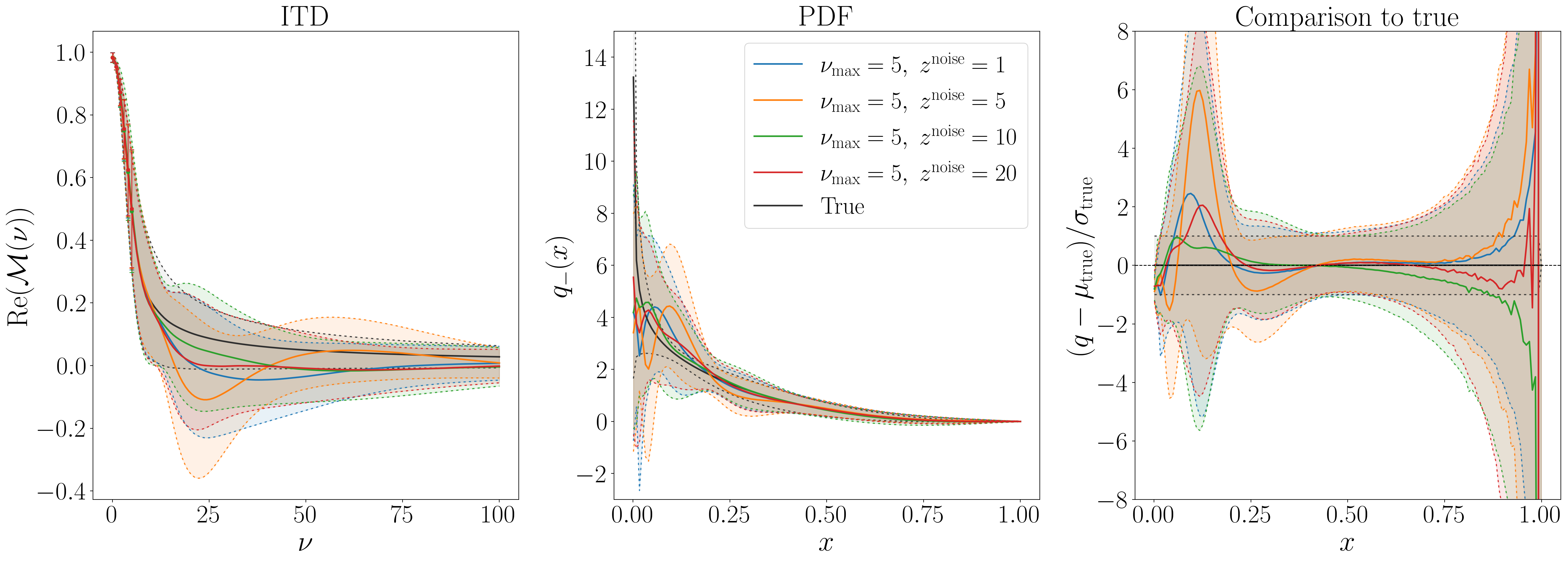}
  \includegraphics[width=0.95\linewidth]{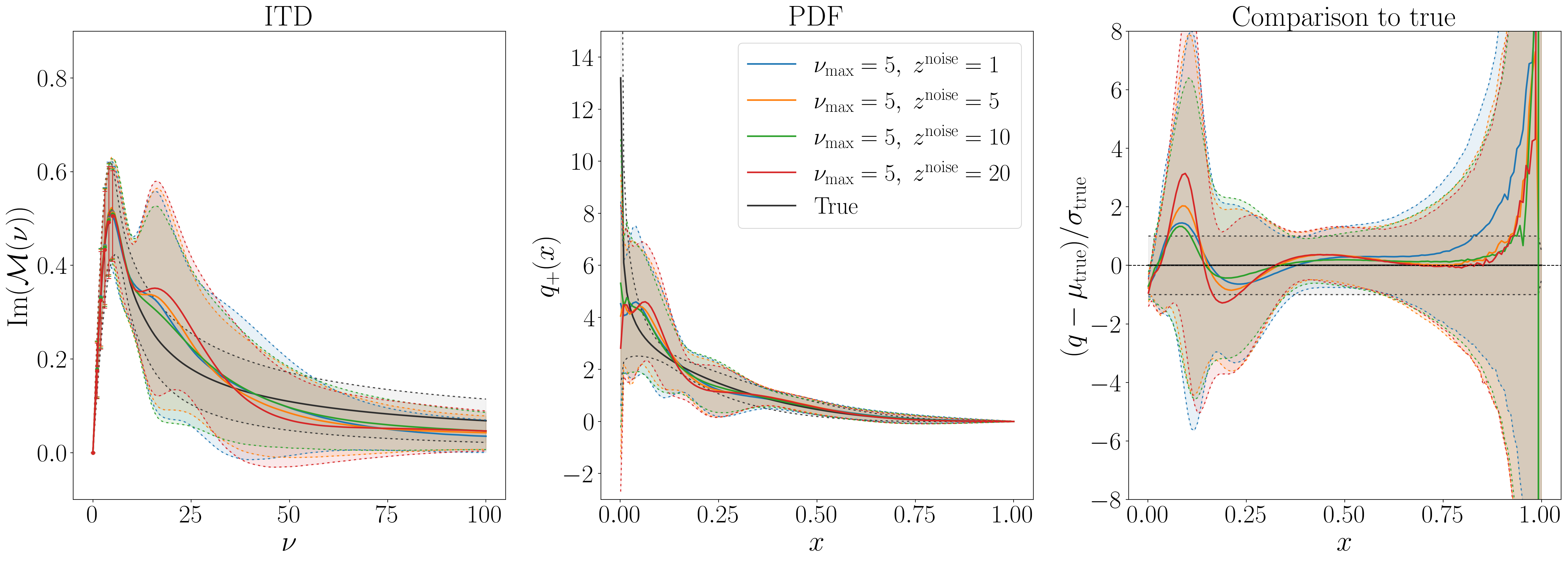}
  \caption{Comparison of different reconstructions of the PDF varying the noise dimensions.}
  \label{fig:noise_inn}
\end{figure}

\begin{figure}[t]
  \centering
  \includegraphics[width=0.95\linewidth]{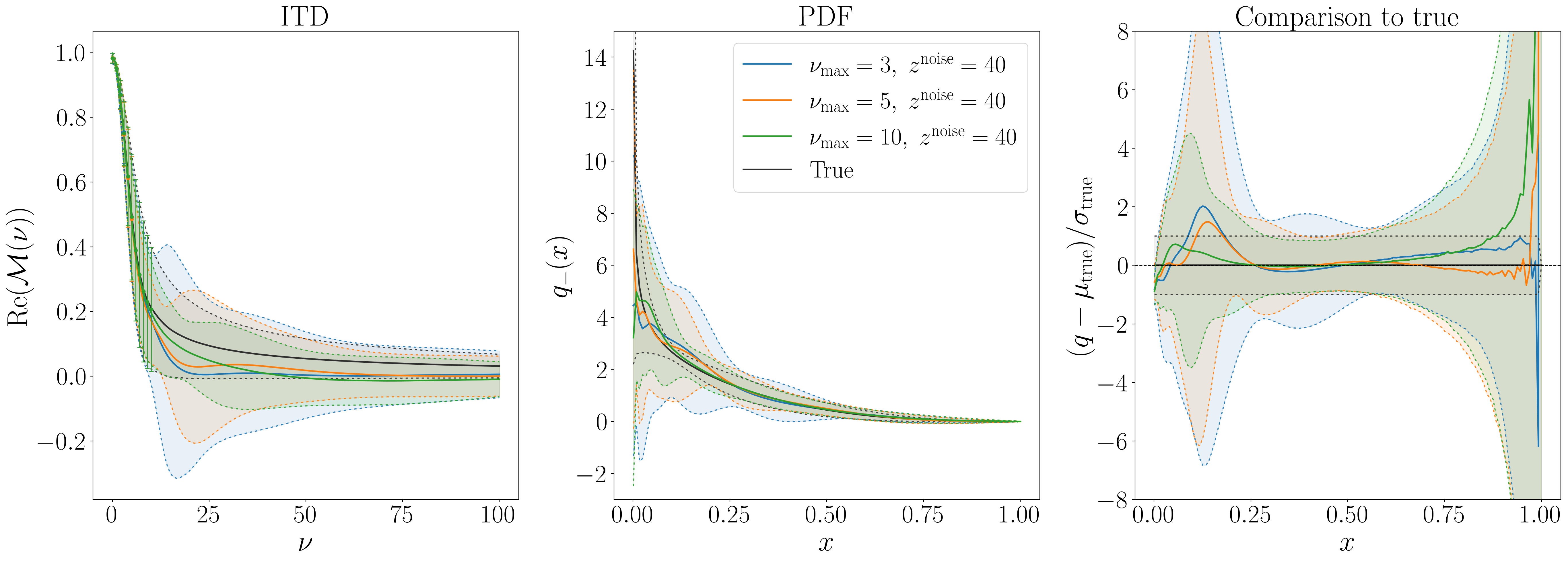}
  \includegraphics[width=0.95\linewidth]{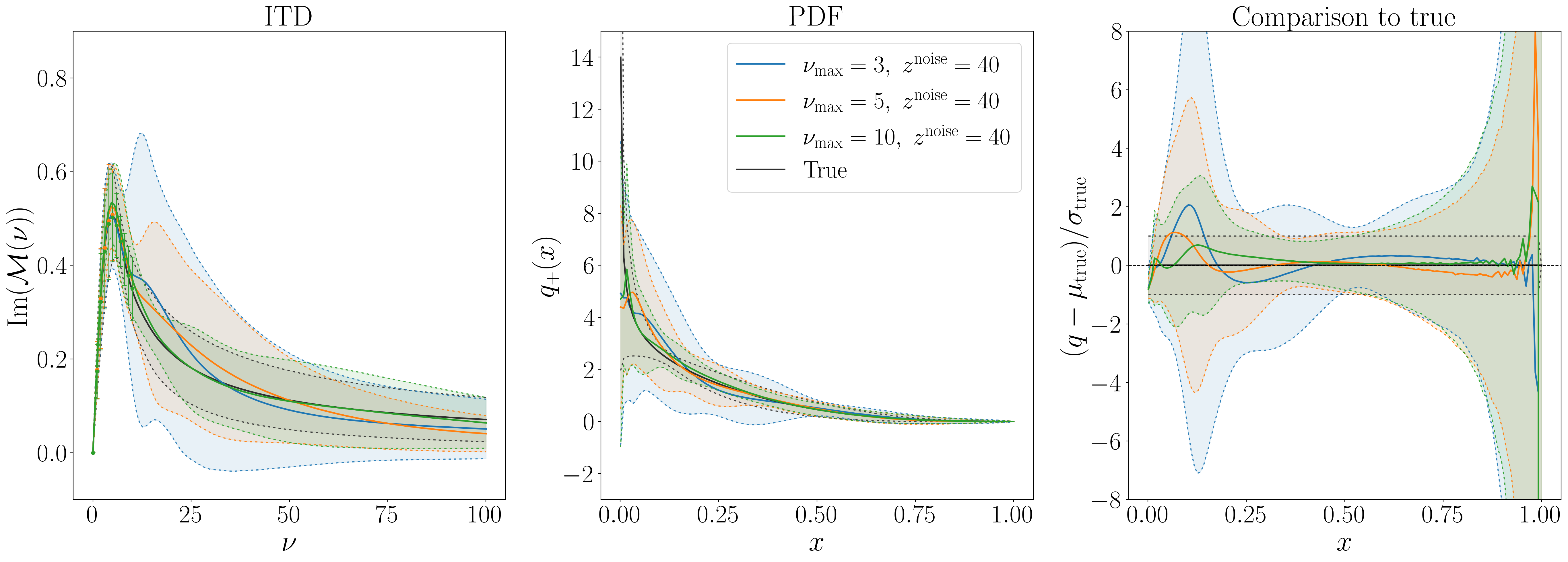}
  \caption{Comparison of different reconstructions of the PDF varying the ITD data available.}
  \label{fig:pdf_nuranges}
\end{figure}

A comparison across different latent-space dimensions is shown in Fig.~\ref{fig:noise_inn}. In this exploration of the dimensionality of the Gaussian noise, the PDF reconstruction remains consistent in the region close to $1$. However, we observe a dependence on the latent dimension in the extrapolation region at low $x$ and high $\nu$, which is an indicator that the regularization of the inverse problem is controlled by the latent/noise variables.
Additionally, the evaluation across different $\nu$ ranges is shown in Fig.~\ref{fig:pdf_nuranges}. It is evident that the reconstruction improves as more information in the ITD is included as the range of $\nu$ grows. This is a feature present in other approaches like Gaussian process     
 regression, Bayesian inference, and parametric regression. 

\section{Discussion}
On the one hand, the architecture successfully preserves the GP prior constraints, such as normalization and the boundary condition $q(x=1)=0$. On the other hand, we identify two main limitations. First, normalization with respect to the prior distribution constrains the INN to operate only on datasets consistent with the prior ITD distribution. This limitation may be alleviated in future work by training on a broader set of GP prior samples generated from different kernel parameterizations and means, as well as by introducing a global mean and variance scale to normalize the data.

A more detailed analysis of the padding and noise dimensions is needed to characterize their impact on the PDF and ITD regions and to identify an optimal configuration. The choice of noise dimensionality is closely related to the criterion employed in Gaussian process regression. Nevertheless, the present approach provides a learned alternative that leads to similar conclusions and serves as a proof of concept for future studies.
For instance, the method can be generalized to reconstruct the dependence on $z$ using real lattice data. It can also be extended to incorporate Mellin moment measurements. Furthermore, for higher-dimensional quantities, such as generalized parton distribution functions, this approach can be naturally extended to those settings.

\section{Conclusion}

We have presented an alternative approach to reconstruct PDFs from synthetic Ioffe-time distribution (ITD) data. The results demonstrate that the INN produces stable reconstructions and captures the main features of the underlying mapping. In this framework, the latent variables encode the missing degrees of freedom in the inverse problem.

While the method inherits the dependence on the choice of prior and latent dimensionality, it offers a data-driven alternative to existing approaches. This work serves as a proof of concept for combining GP priors with invertible neural networks in lattice QCD reconstructions. Future work will focus on extending the framework to real lattice data, incorporating additional observables like Mellin moments, and exploring more expressive flow architectures.

\begin{ack}
   This work is supported by Jefferson
Science Associates, LLC under U.S. DOE Contract \#DE-AC05-06OR23177.
YC acknowledges support in part from the Southeastern Universities Research Association (SURA) through Grant C2024-FEMT-011-03, awarded by the Center for Nuclear Femtography (CNF) and in part by the Fulbright-Garcia Robles scholarship. 


KO was supported in part by U.S.  DOE grant \mbox{
  \#DE-FG02-04ER41302}.
The authors acknowledge William \& Mary Research Computing for providing computational resources and/or technical support that have contributed to the results reported within this paper. In addition, resources of the National Energy Research Scientific Computing Center (NERSC) through the ERCAP project ``{\tt hadron}'' were used. NERSC is a U.S. Department of Energy Office of Science User Facility located at Lawrence Berkeley National Laboratory, operated under Contract No.~DE-AC02-05CH11231.
\end{ack}

\appendix

\bibliographystyle{unsrtnat}
\bibliography{biblio}

\begin{thebibliography}{18}
\providecommand{\natexlab}[1]{#1}
\providecommand{\url}[1]{\texttt{#1}}
\expandafter\ifx\csname urlstyle\endcsname\relax
  \providecommand{\doi}[1]{doi: #1}\else
  \providecommand{\doi}{doi: \begingroup \urlstyle{rm}\Url}\fi

\bibitem[Ji(2013)]{Ji:2013dva}
Xiangdong Ji.
\newblock {Parton Physics on a Euclidean Lattice}.
\newblock \emph{Phys. Rev. Lett.}, 110:\penalty0 262002, 2013.
\newblock \doi{10.1103/PhysRevLett.110.262002}.

\bibitem[Radyushkin(2017)]{Radyushkin:2017cyf}
A.~V. Radyushkin.
\newblock {Quasi-parton distribution functions, momentum distributions, and pseudo-parton distribution functions}.
\newblock \emph{Phys. Rev. D}, 96\penalty0 (3):\penalty0 034025, 2017.
\newblock \doi{10.1103/PhysRevD.96.034025}.

\bibitem[Rasmussen and Williams(2006)]{Rasmussen2006Gaussian}
Carl~Edward Rasmussen and Christopher K.~I. Williams.
\newblock \emph{Gaussian Processes for Machine Learning}.
\newblock The MIT Press, 2006.

\bibitem[Raissi et~al.(2017)Raissi, Perdikaris, and Karniadakis]{Raissi_2017}
Maziar Raissi, Paris Perdikaris, and George~Em Karniadakis.
\newblock Machine learning of linear differential equations using gaussian processes.
\newblock \emph{Journal of Computational Physics}, 348:\penalty0 683–693, November 2017.
\newblock ISSN 0021-9991.
\newblock \doi{10.1016/j.jcp.2017.07.050}.
\newblock URL \url{http://dx.doi.org/10.1016/j.jcp.2017.07.050}.

\bibitem[Karpie et~al.(2019)Karpie, Orginos, Rothkopf, and Zafeiropoulos]{Karpie_2019}
Joseph Karpie, Kostas Orginos, Alexander Rothkopf, and Savvas Zafeiropoulos.
\newblock Reconstructing parton distribution functions from ioffe time data: from bayesian methods to neural networks.
\newblock \emph{Journal of High Energy Physics}, 2019\penalty0 (4), April 2019.
\newblock ISSN 1029-8479.
\newblock \doi{10.1007/jhep04(2019)057}.
\newblock URL \url{http://dx.doi.org/10.1007/JHEP04(2019)057}.

\bibitem[Del~Debbio et~al.(2021)Del~Debbio, Giani, Karpie, Orginos, Radyushkin, and Zafeiropoulos]{Del_Debbio_2021}
Luigi Del~Debbio, Tommaso Giani, Joseph Karpie, Kostas Orginos, Anatoly Radyushkin, and Savvas Zafeiropoulos.
\newblock Neural-network analysis of parton distribution functions from ioffe-time pseudodistributions.
\newblock \emph{Journal of High Energy Physics}, 2021\penalty0 (2), February 2021.
\newblock ISSN 1029-8479.
\newblock \doi{10.1007/jhep02(2021)138}.
\newblock URL \url{http://dx.doi.org/10.1007/JHEP02(2021)138}.

\bibitem[Cichy et~al.(2019)Cichy, Del~Debbio, and Giani]{Cichy_2019}
Krzysztof Cichy, Luigi Del~Debbio, and Tommaso Giani.
\newblock Parton distributions from lattice data: the nonsinglet case.
\newblock \emph{Journal of High Energy Physics}, 2019\penalty0 (10), October 2019.
\newblock ISSN 1029-8479.
\newblock \doi{10.1007/jhep10(2019)137}.
\newblock URL \url{http://dx.doi.org/10.1007/JHEP10(2019)137}.

\bibitem[Alexandrou et~al.(2020)Alexandrou, Iannelli, Jansen, and Manigrasso]{Alexandrou_2020}
Constantia Alexandrou, Giovanni Iannelli, Karl Jansen, and Floriano Manigrasso.
\newblock Parton distribution functions from lattice qcd using bayes-gauss-fourier transforms.
\newblock \emph{Physical Review D}, 102\penalty0 (9), November 2020.
\newblock ISSN 2470-0029.
\newblock \doi{10.1103/physrevd.102.094508}.
\newblock URL \url{http://dx.doi.org/10.1103/PhysRevD.102.094508}.

\bibitem[Candido et~al.(2024)Candido, Debbio, Giani, and Petrillo]{candido2024bayes}
Alessandro Candido, Luigi~Del Debbio, Tommaso Giani, and Giacomo Petrillo.
\newblock Bayesian inference with gaussian processes for the determination of parton distribution functions, 2024.
\newblock URL \url{https://arxiv.org/abs/2404.07573}.

\bibitem[Medrano et~al.(2026)Medrano, Dutrieux, Karpie, Orginos, and Zafeiropoulos]{Medrano:2025}
Yamil~Cahuana Medrano, Hervé Dutrieux, Joseph Karpie, Kostas Orginos, and Savvas Zafeiropoulos.
\newblock Gaussian processes for inferring parton distributions, 2026.
\newblock URL \url{https://arxiv.org/abs/2510.21041}.

\bibitem[Dutrieux et~al.(2025)Dutrieux, Karpie, Orginos, and Zafeiropoulos]{Dutrieux:2024rem}
Herv{\'e} Dutrieux, Joseph Karpie, Kostas Orginos, and Savvas Zafeiropoulos.
\newblock {Simple nonparametric reconstruction of parton distributions from limited Fourier information}.
\newblock \emph{Phys. Rev. D}, 111\penalty0 (3):\penalty0 034515, 2025.
\newblock \doi{10.1103/PhysRevD.111.034515}.

\bibitem[Ardizzone et~al.(2019)Ardizzone, Kruse, Wirkert, Rahner, Pellegrini, Klessen, Maier-Hein, Rother, and Köthe]{ardizzone2019analyzinginverseproblemsinvertible}
Lynton Ardizzone, Jakob Kruse, Sebastian Wirkert, Daniel Rahner, Eric~W. Pellegrini, Ralf~S. Klessen, Lena Maier-Hein, Carsten Rother, and Ullrich Köthe.
\newblock Analyzing inverse problems with invertible neural networks, 2019.
\newblock URL \url{https://arxiv.org/abs/1808.04730}.

\bibitem[Papamakarios et~al.(2021)Papamakarios, Nalisnick, Rezende, Mohamed, and Lakshminarayanan]{papamakarios2021norm}
George Papamakarios, Eric Nalisnick, Danilo~Jimenez Rezende, Shakir Mohamed, and Balaji Lakshminarayanan.
\newblock Normalizing flows for probabilistic modeling and inference, 2021.
\newblock URL \url{https://arxiv.org/abs/1912.02762}.

\bibitem[Dinh et~al.(2017)Dinh, Sohl-Dickstein, and Bengio]{dinh2017densityestimationusingreal}
Laurent Dinh, Jascha Sohl-Dickstein, and Samy Bengio.
\newblock Density estimation using real nvp, 2017.
\newblock URL \url{https://arxiv.org/abs/1605.08803}.

\bibitem[Bishop and Bishop(2023)]{bishop2023learning}
Christopher~Michael Bishop and Hugh Bishop.
\newblock \emph{Deep Learning - Foundations and Concepts}.
\newblock 1 edition, 2023.
\newblock ISBN 978-3-031-45468-4.
\newblock \doi{https://doi.org/10.1007/978-3-031-45468-4}.

\bibitem[Mehraban and Pichler(2025)]{mMMD}
Zahra Mehraban and Alois Pichler.
\newblock Quantization of probability measures in maximum~mean~discrepancy distance, 2025.
\newblock URL \url{https://arxiv.org/abs/2503.11868}.

\bibitem[Kingma and Ba(2017)]{kingma2017adammethodstochasticoptimization}
Diederik~P. Kingma and Jimmy Ba.
\newblock Adam: A method for stochastic optimization, 2017.
\newblock URL \url{https://arxiv.org/abs/1412.6980}.

\bibitem[Loshchilov and Hutter(2017)]{loshchilov2017sgdrstochasticgradientdescent}
Ilya Loshchilov and Frank Hutter.
\newblock Sgdr: Stochastic gradient descent with warm restarts, 2017.
\newblock URL \url{https://arxiv.org/abs/1608.03983}.

\end{thebibliography}

\end{document}